%% file: 0paper.tex
\documentclass[sigconf]{acmart}
\usepackage[T1]{fontenc}

\renewcommand\footnotetextcopyrightpermission[1]{} 
\setcopyright{none}

\usepackage{tikz}
\usepackage{dblfloatfix}
\usepackage{enumitem}
\usepackage{xspace}
\usepackage{balance}
\usepackage{multirow}
\usepackage{subfigure}
\usepackage{color, colortbl}
\usetikzlibrary{shapes,arrows}
\usetikzlibrary{backgrounds, positioning, fit}
\definecolor{LightCyan}{rgb}{0.88,1,1}
\usetikzlibrary{shapes,arrows}
\usepackage{pgfplots}
\usepackage{amsthm}
\newtheorem*{remark}{Remark}
\makeatletter
\tikzset{
    database/.style={
        path picture={
            \draw (0, 1.5*\database@segmentheight) circle [x radius=\database@radius,y radius=\database@aspectratio*\database@radius];
            \draw (-\database@radius, 0.5*\database@segmentheight) arc [start angle=180,end angle=360,x radius=\database@radius, y radius=\database@aspectratio*\database@radius];
            \draw (-\database@radius,-0.5*\database@segmentheight) arc [start angle=180,end angle=360,x radius=\database@radius, y radius=\database@aspectratio*\database@radius];
            \draw (-\database@radius,1.5*\database@segmentheight) -- ++(0,-3*\database@segmentheight) arc [start angle=180,end angle=360,x radius=\database@radius, y radius=\database@aspectratio*\database@radius] -- ++(0,3*\database@segmentheight);
        },
        minimum width=2*\database@radius + \pgflinewidth,
        minimum height=3*\database@segmentheight + 2*\database@aspectratio*\database@radius + \pgflinewidth,
    },
    database segment height/.store in=\database@segmentheight,
    database radius/.store in=\database@radius,
    database aspect ratio/.store in=\database@aspectratio,
    database segment height=0.1cm,
    database radius=0.25cm,
    database aspect ratio=0.35,
}
\makeatother

\newcommand{\pr}{\textsc{PrecRec}\xspace}
\newcommand{\deeplift}{\textsc{DeepLift}\xspace}
\newcommand{\intgrad}{\textsc{IntegratedGradients}\xspace}
\newcommand{\gshap}{\textsc{GradientSHAP}\xspace}
\newcommand{\smgrad}{\textsc{SmoothGrad}\xspace}
\newcommand{\race}{\textsc{Race}\xspace}
\newcommand{\sex}{\textsc{Sex}\xspace}
\newcommand{\mb}{\protect{$\mathcal{M}$}\xspace}
\newcommand{\adv}{\protect{$\mathcal{A}dv$}\xspace}

\begin{document}

\title{Inferring Sensitive Attributes from Model Explanations}

\author{Vasisht Duddu}
\affiliation{%
 \institution{University of Waterloo}
 \city{Waterloo}
 \country{Canada}
}
\email{vasisht.duddu@uwaterloo.ca}

\author{Antoine Boutet}
\affiliation{%
 \institution{Univ Lyon, INSA Lyon, Inria, CITI}
 \city{Lyon}
 \country{France}
}
\email{antoine.boutet@insa-lyon.fr}

\begin{abstract}
Model explanations provide transparency into a trained machine learning model's blackbox behavior to a model builder. They indicate the influence of different input attributes to its corresponding model prediction. The dependency of explanations on input raises privacy concerns for sensitive user data. However, current literature has limited discussion on privacy risks of model explanations.

We focus on the specific privacy risk of \textit{attribute inference attack} wherein an adversary infers sensitive attributes of an input (e.g., \race and \sex) given its model explanations. 
We design the first attribute inference attack against model explanations in two threat models where model builder either (a) includes the sensitive attributes in training data and input or (b) censors the sensitive attributes by not including them in the training data and input.

We evaluate our proposed attack on four benchmark datasets and four state-of-the-art algorithms. 
We show that an adversary can successfully infer the value of sensitive attributes from explanations in both the threat models accurately. Moreover, the attack is successful even by exploiting only the explanations corresponding to sensitive attributes. These suggest that our attack is effective against explanations and poses a practical threat to data privacy.

On combining the model predictions (an attack surface exploited by prior attacks) with explanations, we note that the attack success does not improve. Additionally, the attack success on exploiting model explanations is better compared to exploiting only model predictions. These suggest that model explanations are a strong attack surface to exploit for an adversary.
\end{abstract}




\maketitle
\pagestyle{plain} 

\input{1introduction}
\input{2background}
\input{3problem}
\input{4approach}
\input{5setup}
\input{7results_tm1}
\input{8results_tm2}
\input{9predatt_compare}
\input{10related}

\input{11conclusions}

\bibliographystyle{ACM-Reference-Format}
\balance
\bibliography{paper}

\end{document}

%% file: 1introduction.tex
\section{Introduction}\label{sec:introduction}

Machine Learning (ML) models are used for high-stakes decision making for several real-world applications. For instance, these models assist decision makers such as doctors and judges in healthcare and criminal justice~\cite{rudin2019stop}. However, the model's high complexity makes it difficult for human interpretation into the decision making process. This creates the need for \textit{transparency} into the model behaviour.
Model explanations release additional information to explain the behaviour of complex ML models. Specifically, attribute based model explanations explain the model's prediction on an input by releasing the influence of different input attributes responsible for the prediction~\cite{deeplift,deeplift2,gradshap,intgrad,smoothgrad}. 

Some of the input attributes can be sensitive (e.g., \race and \sex). This raises the data privacy concerns when an adversary (\adv) can leverage model explanations as an attack surface. For instance, Shokri et al.~\cite{shokri2021aies} show that explanations can be exploited for membership inference (i.e., inferring whether input record was part of training data) and data reconstruction.
Additionally, releasing model explanations could leak the values of sensitive attributes which is a privacy risk, not considered in literature. For instance, consider the setting where an ML model is trained to predict the likelihood that a criminal will re-offend as an aid to judges in a court. In addition to output predictions, the model reveals explanations on why it made the prediction on that input. Attribute inference attacks could reveal \race and \sex from model explanations which individual prefers to keep their private to avoid biased decisions. 

However, this quantification of privacy risk of model explanations to \textit{attribute inference attacks} is lacking in current literature. An analysis of this trade-off between privacy and transparency is necessary so that a model builder (\mb) can make appropriate choices to train ML models for high-stakes applications. In this work, we ask the following research question: \textit{can an \adv exploit model explanations to infer sensitive attributes of individual data records?} 
We design the \textit{first attribute inference attack} to infer sensitive attributes from model explanations in two threat models: 
\begin{enumerate}[label=\textbf{TM\arabic*},leftmargin=*]
    \item \label{threatmodel1} Sensitive attributes are included in the training dataset and the input (following prior work~\cite{fredrikson1,fredrikson2}) and \adv only sees the output predictions but not their inputs. \adv has no control over passing the inputs but has to infer sensitive attributes from only the observed predictions.
    \item \label{threatmodel2} Sensitive attributes are not included in training data or input (censored by \mb for privacy). This corresponds to real-world application such as ML as a Service (MLaaS). 
\end{enumerate}

\noindent In this work, we claim the following main contributions.
\begin{enumerate}[leftmargin=*]
    \item We design the \textbf{first} attribute inference attack, to infer sensitive attributes, e.g., \race and \sex, of the data records from corresponding model explanations. \adv trains an ML attack model to map model explanations to sensitive attributes. We additionally calibrate the threshold over the attack model's predictions to increase \adv's power  (Section~\ref{sec:approach}).

    \item In~\ref{threatmodel1}, we show that our attack successfully infers the sensitive attributes from model explanations (Section~\ref{sec:tm1results}). On evaluating across four benchmark datasets and four model explanations, we note: 
    \begin{itemize}
    \item a high F1-score of 0.92 $\pm$ 0.07 (\race) and 0.88 $\pm$ 0.11 (\sex) using entire model explanations corresponding to both sensitive and non-sensitive attributes (Section~\ref{subsec:worstcase}).
    \item a high F1-score of 0.90 $\pm$ 0.10 (\race) and 0.83 $\pm$ 0.09 (\sex) using model explanations corresponding to only sensitive attribute  (Section~\ref{subsec:correlation}).
    \end{itemize}
    
    \item In~\ref{threatmodel2}, despite censoring the sensitive attributes, we show that our attack can successfully infer them using model explanations of other non-sensitive attributes (Section~\ref{sec:tm2results}). On evaluating across four benchmark datasets and four model explanations, we note: 
    \begin{itemize}
        \item a high F1-score of 0.83 $\pm$ 0.12 (\race) and 0.77 $\pm$ 0.09 (\sex)  (Section~\ref{subsec:explonly}).
        \item that on combining model explanations with model predictions, attack success does not improve. Hence, model explanations are a strong attack surface for \adv to exploit (Section~\ref{subsec:combined}).
    \end{itemize}

    \item In both \ref{threatmodel1} and \ref{threatmodel2}, exploiting model explanations has a higher success than prior state-of-the-art attribute inference attacks which exploit model predictions. This indicates that releasing model explanations increases the attack surface enabling \adv to mount strong attribute inference attacks  (Section~\ref{sec:compare}).
\end{enumerate}

%% file: 2background.tex
\section{Background}\label{sec:background}

Consider a training dataset $\mathcal{D} = \{\mathcal{X},\mathcal{S},\mathcal{Y}\}$ where $\mathcal{X}$ is the space of non-sensitive input attributes, $\mathcal{S}$ is the space of sensitive input attributes, $\mathcal{Y}$ is the space of classification labels. We denote a data record as $(x,y,s)$ with non-sensitive attributes $x$ and sensitive attribute $s$ where $(x, s) \in \mathcal{X} \times \mathcal{S}$ and classification label $y \in \mathcal{Y}$.
ML models learn a function $f_{\theta}: (x \cup s) \rightarrow y$ which maps the input with sensitive and non-sensitive attributes to $y$. Alternatively, the models can be trained without $s$ in the training dataset given by $f_{\theta}: x \rightarrow y$. The models are parameterized by $\theta$ which are iteratively updated to minimize the loss on correctly predicting $x$ or $x \cup s$ as $y$. The model training, hyperparameters selection and deployment to application is done by \mb.

Given these formal notations, we describe the state-of-the-art algorithms for model explanations considered in this work (Section~\ref{back:explanations}) and prior work on attribute inference attacks (Section~\ref{back:aia}).

\subsection{Model Explanations}\label{back:explanations}

Model explanations describe a model's behaviour to \mb on specific inputs. Specifically, attribute based model explanations estimate the influence of input attributes on the model's output prediction. In other words, these explanations assign a score to each attribute in the input point of interest (PoI) which resulted in a particular model prediction.
Formally, for a given PoI $\vec{x} = (x_1, \cdots x_n)$, the model explanations $\phi(\vec{x})$ outputs a vector indicating the importance of different attributes influential in the model's prediction of $\vec{x}$.
Here, $\phi(\vec{x})$'s attribution of the prediction at input PoI $\vec{x}$ relative to a baseline input $\vec{x'}$ is a vector
$\phi_{\vec{x'}}(\vec{x}) = (\phi_1,\cdots,\phi_n)$.

We consider two types of attribute based explanation algorithms: (a) backpropagation-based explanations (\intgrad and \deeplift) and (b) perturbation-based explanations (\gshap and \smgrad).

\noindent\textbf{\underline{Gradient-based Explanations}} compute gradients using backpropagation to estimate the influence of attributes to predictions.

\begin{itemize}[leftmargin=*]
     
\item \noindent\textbf{\intgrad}~\cite{intgrad} computes the integration of gradients with respect to inputs by considering a straight line path from the baseline $\vec{x'}$ to the PoI $\vec{x}$. This integration across the $i^{th}$ dimension can be computed as: $\phi_{\intgrad_i}(\vec{x}) = (\vec{x} - \vec{x'}) \times \int_{\alpha=0}^{1} \frac{\partial f_{\theta}(\vec{x'}+ \alpha(\vec{x} - \vec{x'}))}{\partial x_i} \,d\alpha$.
Here, $\frac{\partial f_{\theta}(x)}{\partial x_i}$ indicates the gradient computed using the model $f_{\theta}$ over the input $x$ across the $i^{th}$ dimension.

\item \noindent\textbf{\deeplift}~\cite{deeplift,deeplift2} estimates the contribution of specific neurons using the difference in output with respect to a baseline output. It assignes scores as $\phi_{\deeplift}(\vec{x}) = m_{\Delta \vec{x} \Delta t}\frac{C_{\Delta \vec{x}\Delta t}}{\Delta \vec{x}}$ where $x$ is a given input neuron, $\Delta x$ is the difference from baseline, $t$ is target neuron and its output difference from baseline is given as $\Delta t$. The multiplier captures the contribution of $\Delta x$ to $\Delta t$ and is similar to partial derivative but over finite differences instead of infinitesimal differences. 
\end{itemize}

\noindent\textbf{\underline{Perturbation-based Explanations}} add noise to data records or remove some attributes to see the impact on the model utility.

\begin{itemize}[leftmargin=*]

\item \noindent\textbf{\gshap}~\cite{gradshap} computes the Shapley values and adds Gaussian noise to each input PoI by sampling multiple times and selects a random input $x$ along the path between baseline and input. The gradient of outputs with respect to those selected random points are then computed. 
In other words, the final attributes are computed as the expected value over the product of the gradient and the difference in input PoI to the baseline: $\frac{\partial f_{\theta}(x)}{\partial x} \times (\vec{x} - \vec{x'})$.

\item \noindent\textbf{\smgrad}~\cite{smoothgrad} samples random inputs in a neighborhood of PoI $\vec{x}$ by adding Gaussian noise to the PoI. Then it averages the resulting sensitivity maps (i.e., derivative of model predictive with respect to input) corresponding to the $n$ noisy neighbour records $\phi_{\smgrad}(\vec{x}) = \frac{1}{n}\sum_{1}^{n}\frac{\partial f_{\theta}(\vec{x} + \mathcal{N}(0,\sigma^2))}{\vec{x}}$.

\end{itemize}

A natural choice for baseline $\vec{x'}$ to compute model explanations is where the prediction is unbiased~\cite{intgrad}. In all the cases, we use the mean vector over the inputs as our baseline. Additionally, each model explanation algorithm also outputs a convergence delta, $\delta$ where the lower the absolute value of the convergence delta the better is the approximation (i.e., low error). We append $\delta$ with $\phi()$ to obtain the final attack vector. We abuse the notation to refer the appended vector as $\phi()$.

\subsection{Attribute Inference Attacks}\label{back:aia}

Attribute inference attacks aim to infer $s$ (e.g., $s=1$ for males and $s=0$ for females) for an individual data record. \adv exploits observable information (i.e., model predictions or explanations in our case) to infer unobservable information (i.e., $s$). This attack is different from property inference attacks proposed in literature which aim to infer global properties of dataset (e.g., inferring the ratio of males to female attributes on which the model was trained on)~\cite{Melis2019ExploitingUF,propinf1,tople21usenix}.

Several prior work have proposed attribute inference attacks against ML models using the model's output predictions~\cite{fredrikson1,fredrikson2,Song2020Overlearning,yeom,Mahajan2020DoesLS,malekzadeh2021honest}. Fredrikson et al.~\cite{fredrikson1,fredrikson2} propose an attribute inference attack where \adv infers $s$ using the knowledge of both $x$ and $f_{\theta}(x \cup s)$. However, this assumption of \adv's knowledge is strong. Mahajan et al.~\cite{Mahajan2020DoesLS} and Song et al.~\cite{Song2020Overlearning} proposed an attack where an ML attack model was trained to infer $s$ using only model prediction. This attack exploits the distinguishability in predictions conditioned on different values of $s$. However, the attack model performs poorly for imbalanced dataset since the default threshold of 0.5 for estimating the value of $s$ is incorrect for skewed prediction distribution. To address this, Aalmoes et al.~\cite{aalmoes2022dikaios} proposed an attack which accounts for this skewness of attack model's predictions. They select a threshold over attack model's predictions which maximizes attack F1-Score on an auxiliary dataset known to \adv. That threshold is used over attack model's predictions to infer $s$ for target data records.

%% file: 3problem.tex
\section{Problem Statement}\label{sec:problem}

Our goal is to evaluate the privacy risks of model explanations to attribute inference attacks and hence study the trade-offs between privacy and transparency.
We consider the following setting: target ML model $f_{target}$ is trained and deployed on the Cloud by \mb within MLaaS paradigm. Given a POI $\vec{x}$, we assume that $f_{target}$ can output both the model prediction ($f_{target}(\vec{x})$) and corresponding explanations on that input ($\phi(\vec{x})$). $\phi()$ are required to be released by AI regulations to ensure trustworthy computation~\cite{ec2019ethics,dpia,nist,ico,whitehouse}.

Given that model explanations measure the influence of individual attributes in the input to the model's prediction, it is natural to ask, \textit{given access to $\phi()$, can \adv infer $s$?} This study is currently lacking in literature. We describe three main requirements for the design of an effective attack:
\begin{enumerate}[label=\textbf{AR\arabic*},leftmargin=*]
    \item \label{attreq1} Attack should operate in a \textbf{blackbox threat model}, where \adv sends an input and obtains an output via an API from a MLaaS service provider. \adv does not have access to $f_{target}$'s internal parameters or architecture.
    \item \label{attreq2} Attack should be \textbf{practical}, i.e., uses model observables ($\phi()$ or $f_{target}()$) to infer unobservables ($s$).
    \item \label{attreq3} Attack should \textbf{account for class imbalance in $s$}. In all practical applications, $s$ is imbalanced which skews the predictions of $f_{adv}$ lowering the \adv's attack success to correctly infer $s$. 
    \item \label{attreq4} Attack should be \textbf{applicable to model explanations}, i.e., exploits $\phi()$ to infer the values of $s$. 
\end{enumerate}

Prior attacks exploit the distinguishability in $f_{target}()$ given different values of $s$~\cite{fredrikson1,fredrikson2,yeom,Song2020Overlearning,Mahajan2020DoesLS,aalmoes2022dikaios}. Fredrikson et al.~\cite{fredrikson1,fredrikson2} and Yeom et al.~\cite{yeom} attacks have strong assumptions about \adv knowledge, such as knowledge of $x$ in addition to $f_{target}()$ (violating requirement~\ref{attreq2}). Alternatively, Song et al.~\cite{Song2020Overlearning} and Mahajan et al.~\cite{Mahajan2020DoesLS} do not account for the class imbalance in $s$ (violating requirement~\ref{attreq3}). Aalmoes et al.~\cite{aalmoes2022dikaios} use threshold calibration to improve the attack success but they exploit $f_{target}()$ and not $\phi()$ (violating \ref{attreq4}).

\subsection{Threat Model and Attack Methodology}\label{sec:threatmodel}

We discuss two threat models \ref{threatmodel1} and \ref{threatmodel2} along with the assumptions about \adv's knowledge and attack methodology.

\begin{itemize}[leftmargin=*]
\item \noindent\textbf{\ref{threatmodel1} (w/ $s$ in $\mathcal{D}$):} We assume $s$ is included in both $\mathcal{D}$ and input (i.e., $x \cup s$). Hence, $\phi(x \cup s)$ are released as part of the API and \adv can obtain $\phi(s)$ along with $\phi(x)$. This is when the \adv is monitoring the outputs from the model. Here, \adv cannot choose inputs to send to $f_{target}$ (as it already includes $s$)\footnote{Despite this, this is seen in several prior attribute inference attacks~\cite{fredrikson1,fredrikson2,Song2020Overlearning,yeom,Mahajan2020DoesLS}.} but can only observe $f_{target}(x \cup s)$ and $\phi(x \cup s)$ for some arbitrary inputs. Given only $\phi(x \cup s)$, \adv aims to infer $s$ using an attack ML model $f_{adv}: \phi(x \cup s) \rightarrow s$. Here, \adv can also attack different model explanations: $f_{adv}: \phi(x) \rightarrow s$ and $f_{adv}: \phi(s) \rightarrow s$\footnote{$\phi(s)$ and $\phi(x)$ indicates model explanations corresponding to $s$ and $x$ respectively.}.
\input{fig_tm1}

\item \noindent\textbf{\ref{threatmodel2} (w/o $s$ in $\mathcal{D}$):} We assume $s$ is \textit{not} included in $\mathcal{D}$ and input (i.e., $x$). \adv has blackbox access to $f_{target}$ and pass an input $x$ and obtain access to both $\phi(x)$ and $f_{target}(x)$.  Unlike \ref{threatmodel1}, \adv can choose the input to pass to the model. This is the worst case for \adv where $s$ is censored by \mb for privacy making this threat model more practical. Given $\phi(x)$, \adv aims to infer $s$ using an attack ML model $f_{adv}: \phi(x) \rightarrow s$. 
\input{fig_tm2}
\end{itemize}

Figure~\ref{fig:tm1} and \ref{fig:tm2} where \colorbox{red!25}{red} indicates components accessible to \adv.
In both \ref{threatmodel1} and \ref{threatmodel2}, we assume that \adv has additional auxiliary dataset $\mathcal{D}_{aux}$ which is drawn from the same distribution as $\mathcal{D}$ and includes data records $(x,s,y)$ containing non-sensitive and sensitive attributes with corresponding label. This assumptions is inline with all the prior attribute inference attacks proposed in literature~\cite{fredrikson1,fredrikson2,Song2020Overlearning,yeom,Mahajan2020DoesLS}.
$\mathcal{D}_{aux}$ is used to train $f_{adv}$: \adv passes data records $(x,s,y)$ (\ref{threatmodel1}) or $(x,y)$ (\ref{threatmodel2}) to $f_{target}$ and uses the generated model explanations to train $f_{adv}$ by mapping them to $s$ (known to \adv for $\mathcal{D}_{aux}$). This access to $f_{target}$ can be alleviated by training a ``shadow model'' on $\mathcal{D}_{aux}$ to mimic $f_{target}$; and use the model explanations from ``shadow model'' to train $f_{adv}$. We only use predictions from $f_{target}$ to train $f_{adv}$.
Once $f_{adv}$ is trained, the attack is evaluated on target dataset (distinct from $\mathcal{D}_{aux}$).

Moreover, any attack designed within \ref{threatmodel1} and \ref{threatmodel2} are blackbox (satisfy \ref{attreq1}) and practical (satisfy \ref{attreq2}). In both threat models, $\phi()$ is accessible to adversary to exploit and hence satisfies \ref{attreq4}.
Given this, we now have to design an attack which satisfies requirement~\ref{attreq3} to account for class imbalance in $s$ and improve attack success.

%% file: fig_tm1.tex
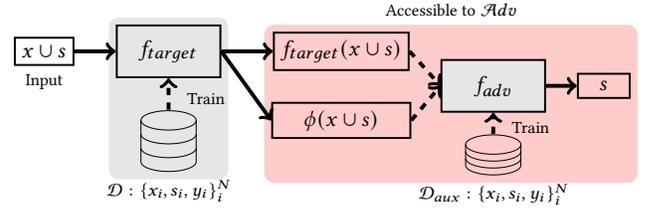
\begin{figure}[h]
\resizebox{.47\textwidth}{!}{%
\begin{tikzpicture}

    \node [rectangle,draw,thick,minimum width=1.5cm, minimum height=0.75cm] (targetmodel) {$f_{target}$};
    
    \node[below of=targetmodel,yshift=-0.3cm,database,database radius=0.4cm,database segment height=0.2cm, label={below:\footnotesize $\mathcal{D}: \{x_i,s_i,y_i\}_{i}^N$}] (trainingdata) {};

    \node [left of=targetmodel,xshift=-0.8cm,rectangle,draw,thick,label={below:\footnotesize Input}] (inputrecord) {$x \cup s$};
    \node [right of=targetmodel,xshift=1.5cm,rectangle,draw,thick] (outputpred) {$f_{target}(x \cup s)$};
    \node [below of=outputpred,minimum width=2cm,rectangle,draw,thick] (explanation) {$\phi(x \cup s)$};

    \begin{scope}[on background layer]
        \node (tm1) [fit=(targetmodel) (trainingdata), fill= gray!20, rounded corners, inner sep=0.1cm, label={above:\footnotesize }] {};
    \end{scope}

    \node [right of=outputpred,rectangle,draw,thick,minimum width=1.5cm, minimum height=0.75cm,xshift=1.2cm,yshift=-0.5cm,fill= gray!20] (attmodel) {$f_{adv}$};

    \node [right of=attmodel,xshift=0.6cm,minimum width=0.75cm,rectangle,draw,thick] (attout) {$s$};
    \node[below of=attmodel,database,database radius=0.4cm,database segment height=0.1cm, label={below:\footnotesize $\mathcal{D}_{aux}: \{x_i,s_i,y_i\}_{i}^N$}] (auxdata) {};

\begin{scope}[on background layer]
    \node (models) [fit=(attmodel) (outputpred) (auxdata) (attout) (explanation), fill= red!20, rounded corners, inner sep=0.1cm, label={above:\footnotesize Accessible to \adv}] {};
\end{scope}

\draw[->,ultra thick] (inputrecord.east) -- node[anchor=south, align=center] {\em\footnotesize } (targetmodel.west);
\draw[->,ultra thick] (targetmodel.east) -- node[anchor=south, align=center] {\em\footnotesize } (outputpred.west);
\draw[->,ultra thick] (targetmodel.east) -- node[anchor=south, align=center] {\em\footnotesize } (explanation.west);

\draw[->,ultra thick, dashed] (outputpred.east) -- node[anchor=south, align=center] {\em\footnotesize } (attmodel.west);
\draw[->,ultra thick, dashed] (explanation.east) -- node[anchor=south, align=center] {\em\footnotesize } (attmodel.west);
\draw[->,ultra thick] (attmodel.east) -- node[anchor=south, align=center] {\em\footnotesize } (attout.west);
\draw[->,ultra thick,dashed] (trainingdata.north) -- node[anchor=south, align=center,label={[yshift=-0.2cm]right:\footnotesize Train}] {\em\footnotesize } (targetmodel.south);
\draw[->,ultra thick,dashed] (auxdata.north) -- node[anchor=south, align=center,label={[yshift=-0.2cm]right:\footnotesize Train}] {\em\footnotesize } (attmodel.south);

\end{tikzpicture}
}
\caption{\ref{threatmodel1} threat model: train $f_{target}$ on training data with $s$ included. \adv only has access to predictions $f_{target}(x \cup s)$ and explanations $\phi_{target}(x \cup s)$ but cannot pass inputs. Attack requires training $f_{adv}$ on $\mathcal{D}_{aux}$ to infer $s$ given $\phi_{target}(x \cup s)$.} 
\label{fig:tm1}
\end{figure}

%% file: fig_tm2.tex
\begin{figure}[h]
\resizebox{.47\textwidth}{!}{%
\begin{tikzpicture}

    \node [rectangle,draw,thick,minimum width=1.5cm, minimum height=0.75cm] (targetmodel) {$f_{target}$};
    
    \node[below of=targetmodel,yshift=-0.3cm,database,database radius=0.4cm,database segment height=0.2cm, label={below:\footnotesize $\mathcal{D}: \{x_i,y_i\}_{i}^N$}] (trainingdata) {};

    \node [left of=targetmodel,minimum width=0.75cm,xshift=-0.8cm,fill=red!20,rectangle,draw,thick,label={below:\footnotesize Input}] (inputrecord) {$x$};
    \node [right of=targetmodel,xshift=1.5cm,rectangle,draw,thick] (outputpred) {$f_{target}(x)$};
    \node [below of=outputpred,minimum width=1.5cm,rectangle,draw,thick] (explanation) {$\phi(x)$};

    \begin{scope}[on background layer]
        \node (tm1) [fit=(targetmodel) (trainingdata), fill= gray!20, rounded corners, inner sep=0.1cm, label={above:\footnotesize }] {};
    \end{scope}

    \node [right of=outputpred,rectangle,draw,thick,minimum width=1.5cm, minimum height=0.75cm,xshift=1.2cm,yshift=-0.5cm,fill= gray!20] (attmodel) {$f_{adv}$};

    \node [right of=attmodel,xshift=0.6cm,minimum width=0.75cm,rectangle,draw,thick] (attout) {$s$};

    \node[below of=attmodel,database,database radius=0.4cm,database segment height=0.1cm, label={below:\footnotesize $\mathcal{D}_{aux}: \{x_i,s_i,y_i\}_{i}^N$}] (auxdata) {};

\begin{scope}[on background layer]
    \node (models) [fit=(attmodel) (outputpred) (attout) (auxdata) (explanation), fill= red!20, rounded corners, inner sep=0.1cm, label={above:\footnotesize Accessible to \adv}] {};
\end{scope}

\draw[->,ultra thick] (inputrecord.east) -- node[anchor=south, align=center] {\em\footnotesize } (targetmodel.west);
\draw[->,ultra thick] (targetmodel.east) -- node[anchor=south, align=center] {\em\footnotesize } (outputpred.west);
\draw[->,ultra thick] (targetmodel.east) -- node[anchor=south, align=center] {\em\footnotesize } (explanation.west);

\draw[->,ultra thick, dashed] (outputpred.east) -- node[anchor=south, align=center] {\em\footnotesize } (attmodel.west);
\draw[->,ultra thick, dashed] (explanation.east) -- node[anchor=south, align=center] {\em\footnotesize } (attmodel.west);
\draw[->,ultra thick] (attmodel.east) -- node[anchor=south, align=center] {\em\footnotesize } (attout.west);
\draw[->,ultra thick,dashed] (trainingdata.north) -- node[anchor=south, align=center,label={[yshift=-0.2cm]right:\footnotesize Train}] {\em\footnotesize } (targetmodel.south);
\draw[->,ultra thick,dashed] (auxdata.north) -- node[anchor=south, align=center,label={[yshift=-0.2cm]right:\footnotesize Train}] {\em\footnotesize } (attmodel.south);

\end{tikzpicture}
}
\caption{\ref{threatmodel2} threat model: train $f_{target}$ on data without $s$. \adv has access to predictions $f_{target}(x)$ and explanations $\phi_{target}(x)$ and choose the inputs to pass (w/o $s$). Attack trains $f_{adv}$ on $\mathcal{D}_{aux}$ to infer $s$ given $\phi_{target}(x)$.} 
\label{fig:tm2}
\end{figure}
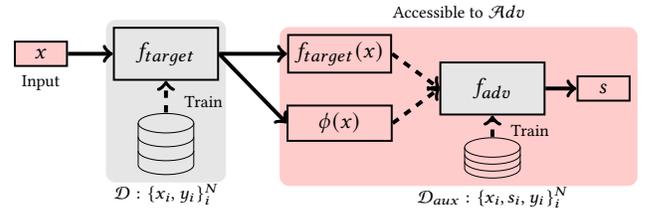

%% file: 4approach.tex
\section{Our Proposed Attack}\label{sec:approach}

Prior attribute inference attacks are directly applicable as they do not satisfy requirements \ref{attreq1}-\ref{attreq4}. We design attribute inference attacks to adapt to $\phi()$ to infer $s$ while calibrating the threshold over $f_{adv}$'s predictions to improve attack success.
Instead of using the default threshold of 0.5, as in prior attacks over model predictions~\cite{Mahajan2020DoesLS,Song2020Overlearning}, we calibrate the threshold over $f_{adv}(\phi())$ to maximize F1-Score.
We compute an optimal threshold $\tau^*$ over the probability $P(s|\phi(x))$, which is the output of $f_{adv}(\phi(x))$, to infer $s$. 
In practice, we use the precision-recall curve which computes precision and recall values for multiple thresholds over $f_{adv}$'s predictions. Then, $\tau^*$ is chosen based on maximum F1-Score and this in-turn improves the precision and recall values. This is effective when there is a moderate to large class imbalance (satisfies~\ref{attreq3}).

\begin{figure*}[h]
    \centering

    \subfigure[CENSUS (\intgrad)]{\includegraphics[width=0.24\textwidth]{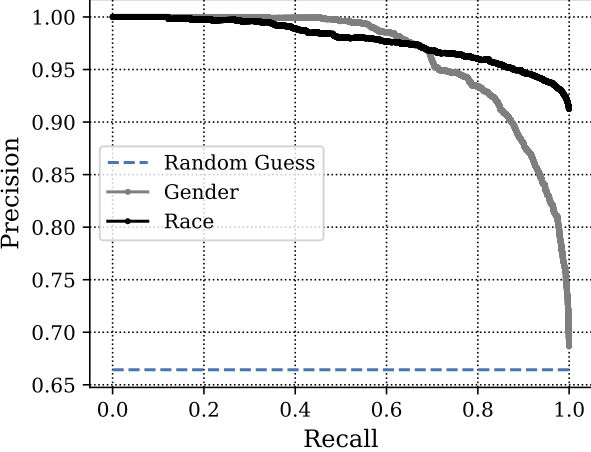}} 
    \subfigure[COMPAS (\intgrad)]{\includegraphics[width=0.24\textwidth]{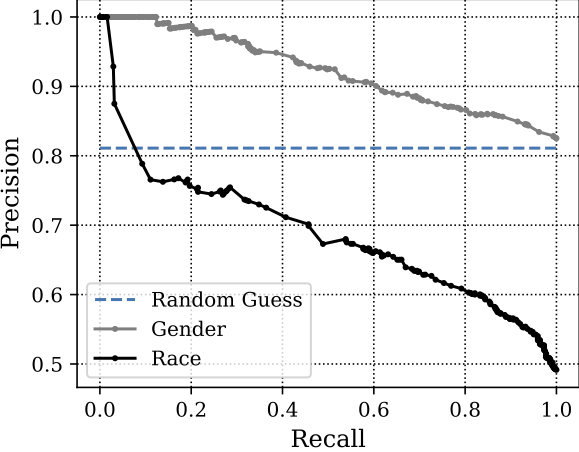}} 
    \subfigure[CREDIT (\intgrad)]{\includegraphics[width=0.24\textwidth]{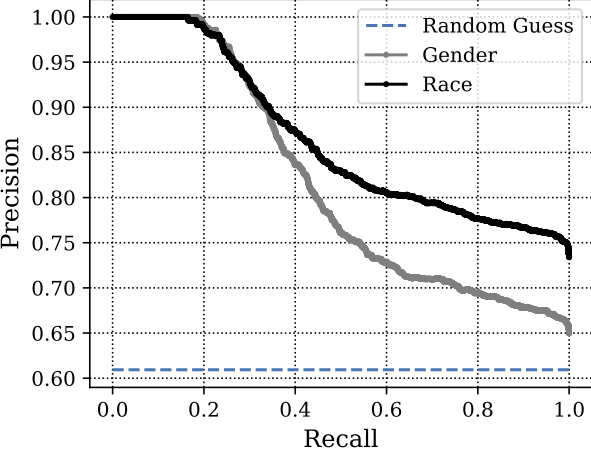}} 
     \subfigure[LAW (\intgrad)]{\includegraphics[width=0.24\textwidth]{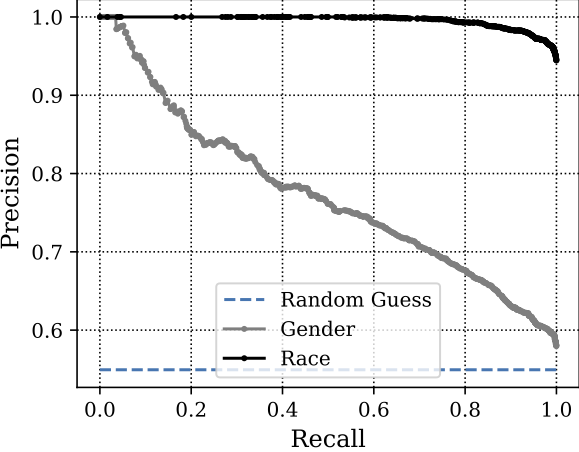}}

    \subfigure[CENSUS (\deeplift)]{\includegraphics[width=0.24\textwidth]{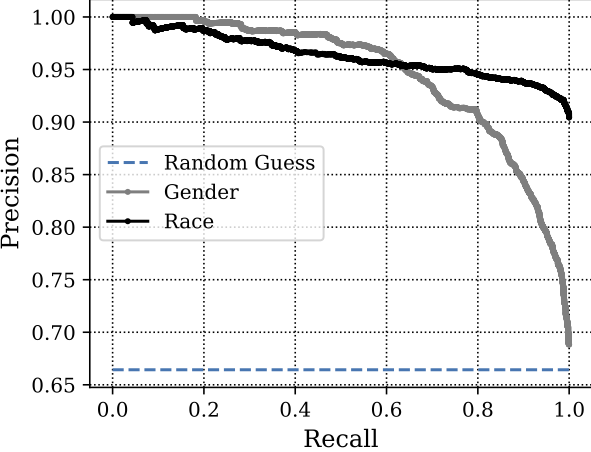}}
    \subfigure[COMPAS (\deeplift)]{\includegraphics[width=0.24\textwidth]{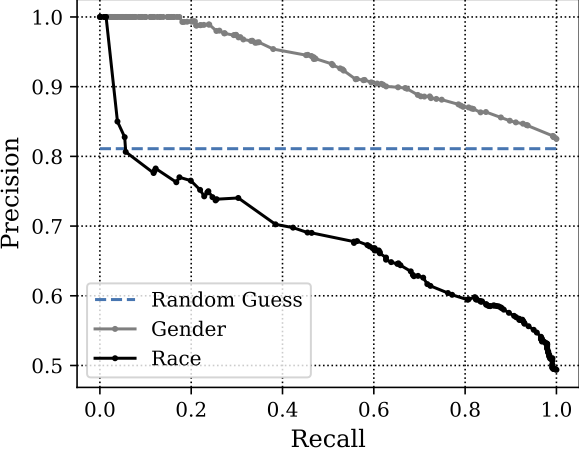}}
    \subfigure[CREDIT (\deeplift)]{\includegraphics[width=0.24\textwidth]{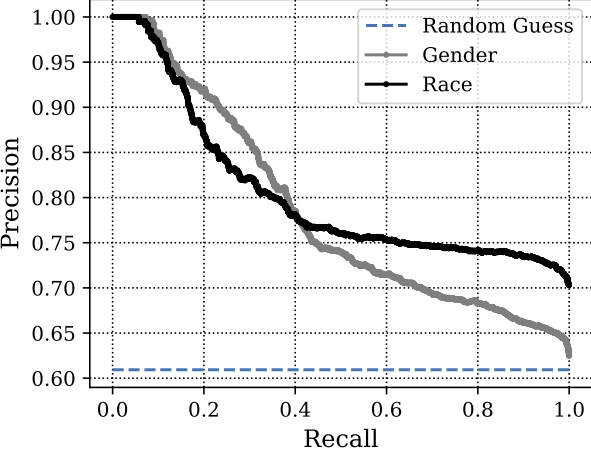}}
    \subfigure[LAW (\deeplift)]{\includegraphics[width=0.24\textwidth]{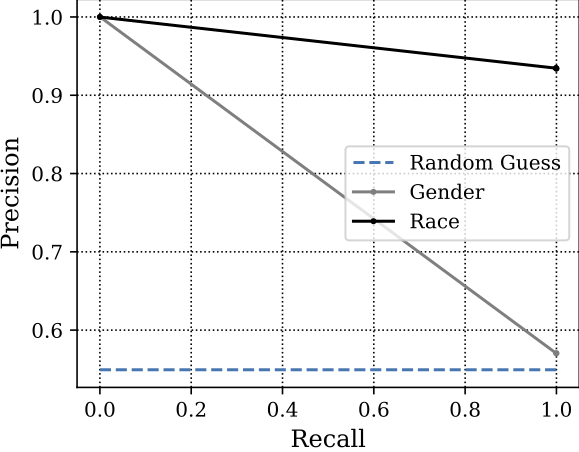}}\\

    \subfigure[CENSUS (\gshap)]{\includegraphics[width=0.24\textwidth]{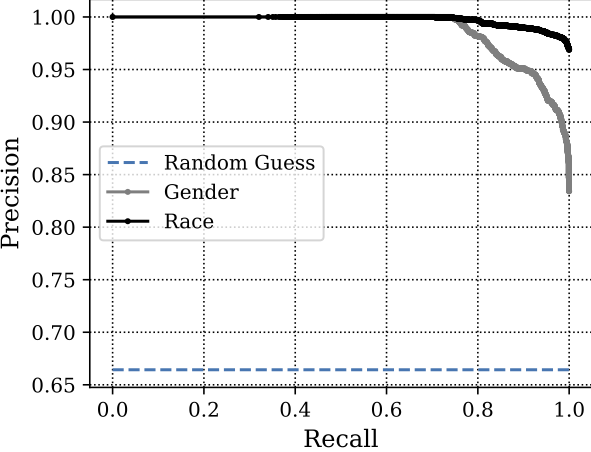}}
    \subfigure[COMPAS (\gshap)]{\includegraphics[width=0.24\textwidth]{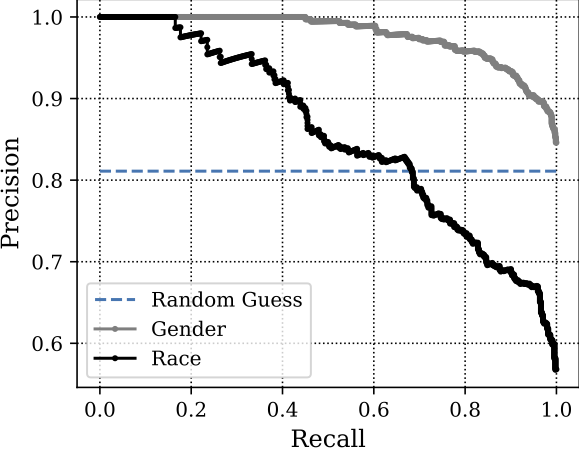}} 
    \subfigure[CREDIT (\gshap)]{\includegraphics[width=0.24\textwidth]{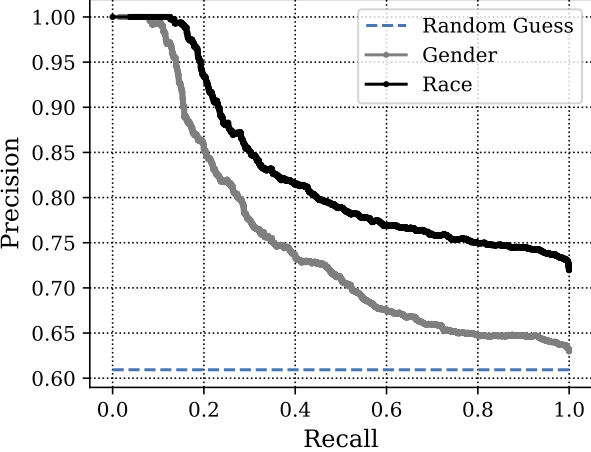}}
    \subfigure[LAW (\gshap)]{\includegraphics[width=0.24\textwidth]{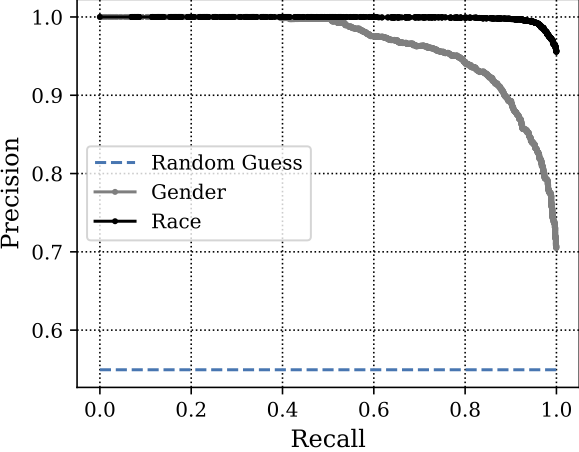}}\\

    \subfigure[CENSUS (\smgrad)]{\includegraphics[width=0.24\textwidth]{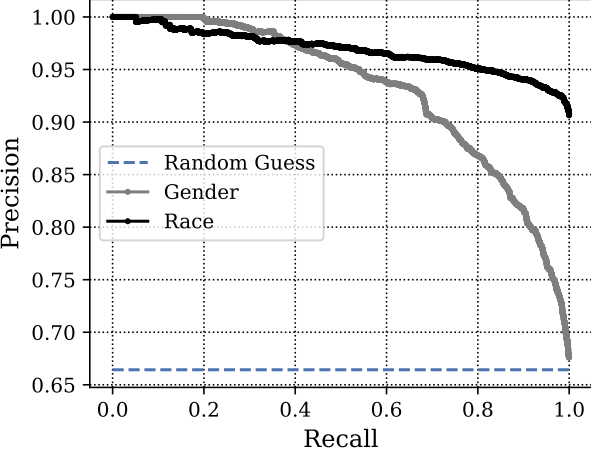}}
    \subfigure[COMPAS (\smgrad)]{\includegraphics[width=0.24\textwidth]{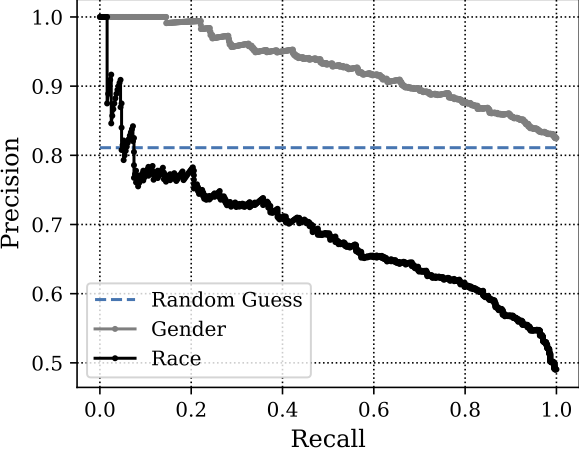}} 
    \subfigure[CREDIT (\smgrad)]{\includegraphics[width=0.24\textwidth]{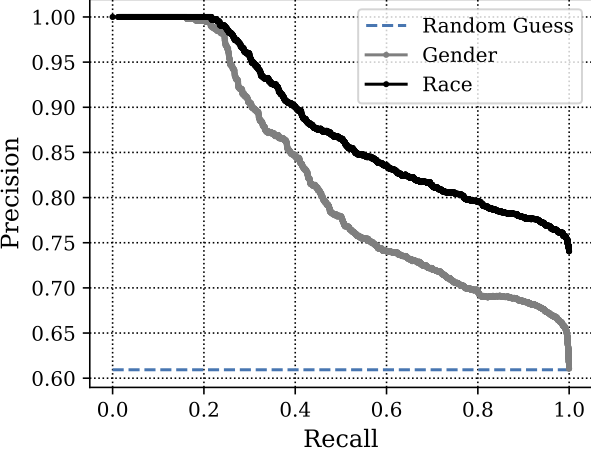}}
    \subfigure[LAW (\smgrad)]{\includegraphics[width=0.24\textwidth]{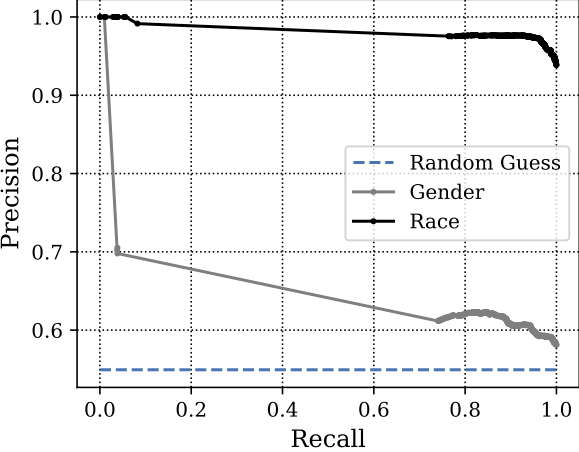}} 

    \caption{Precision-recall curves for finding optimal threshold to improving \adv's success. The precision recall are above random guess which can allow \adv to compute an optimal threshold to improve attack success.}
    \label{fig:prplot}
\end{figure*}

\noindent\textbf{Calibrating the Threshold.} First, as a sanity check, we ensure the precision-recall curves are above random guess baseline. A random guess for precision-recall curve is the horizontal line with the precision value computed over the positive class examples in the dataset. Figure~\ref{fig:prplot} shows the precision-recall curves for $f_{adv}$ on $\mathcal{D}_{aux}$ which is beyond random guess in all cases. 
This indicates the possibility of finding $\tau^*$ to improve \adv's F1-Score.

\setlength\tabcolsep{1.75pt}
\begin{table}[h]
\begin{center}
\caption{$\tau^*$ is different from default threshold of 0.5. ``IG'' is \intgrad, ``DL'' is \deeplift, ``GS'' is \gshap and ``SG'' is \smgrad.\label{tbl:thresholds}}
\begin{tabular}{ |c|c|c|c|c|c|c|c|c|}
 \hline
\rowcolor{LightCyan} &  \multicolumn{4}{|c|}{\textbf{IG}} & \multicolumn{4}{|c|}{\textbf{DL}}\\
  \hline
 & \multicolumn{2}{c|}{w/ S} & \multicolumn{2}{c|}{w/o S} & \multicolumn{2}{c|}{w/ S} & \multicolumn{2}{c|}{w/o S}\\
\textbf{Dataset} & \textbf{\race} & \textbf{\sex} & \textbf{\race} & \textbf{\sex} & \textbf{\race} & \textbf{\sex} & \textbf{\race} & \textbf{\sex}\\
\hline
\textbf{CENSUS} & 0.64 & 0.47 & 0.42 & 0.54 & 0.96 & 0.51 & 0.82 & 0.37  \\
\textbf{COMPAS} & 0.94 & 0.89 & 0.38 & 0.59 & 0.97 & 0.84 & 0.38 & 0.52\\
\textbf{LAW}    & 0.93 & 0.56 & 0.93 & 0.56 & 0.93 & 0.74 & 0.79 & 0.56\\
\textbf{CREDIT} & 0.55 & 0.42 & 0.54 & 0.48 & 0.61 & 0.55 & 0.46 & 0.40\\
 \hline
\rowcolor{LightCyan} & \multicolumn{4}{|c|}{\textbf{GS}} &  \multicolumn{4}{|c|}{\textbf{SG}} \\
  \hline
 & \multicolumn{2}{c|}{w/ S} & \multicolumn{2}{c|}{w/o S}& \multicolumn{2}{c|}{w/ S} & \multicolumn{2}{c|}{w/o S}\\
\textbf{Dataset} & \textbf{\race} & \textbf{\sex} & \textbf{\race} & \textbf{\sex}  & \textbf{\race} & \textbf{\sex} & \textbf{\race} & \textbf{\sex} \\
\hline
\textbf{CENSUS} & 0.77 & 0.26 & 0.55 & 0.47 & 0.68 & 0.49 & 0.51 & 0.50\\
\textbf{COMPAS} & 0.61 & 0.58 & 0.46 & 0.54 & 0.81 & 0.72 & 0.33 & 0.55\\
\textbf{LAW}    & 0.68 & 0.61 & 0.82 & 0.56 & 0.97 & 0.96 & 0.93 & 0.57\\
\textbf{CREDIT} & 0.48 & 0.48 & 0.56 & 0.52 & 0.60 & 0.43 & 0.51 & 0.44 \\
 \hline
 \end{tabular}
\end{center}
\end{table}

Table~\ref{tbl:thresholds} further shows that the resultant $\tau^*$ is indeed different from 0.5 default threshold, indicative of improving attack success. It is important to note that $\tau^*$ is computed on $\mathcal{D}_{aux}$ which might be different from optimal threshold on target dataset which is being attacked. However, this cannot be known before-hand by \adv. This is the best \adv can do before performing the attack in real-world with imbalanced datasets hoping that $\tau^*$ improves attack success.

\noindent\textbf{Why is this a Privacy Risk?} Our threat models are similar to prior attacks~\cite{fredrikson1,fredrikson2,Song2020Overlearning,yeom,Mahajan2020DoesLS}. One can argue that the attack is not actually exploiting the model explanations but using the existing correlations between sensitive and non-sensitive attributes (which \adv could deduce from $\mathcal{D}_{aux}$). In this case, there is no privacy violation.
However, as seen in Table~\ref{tbl:correlation}, Pearson's correlation between $s$ and other attributes is low. Hence, \adv exploits non-trivial information, i.e., information memorized by $f_{target}$ about $s$ which is present in $\phi()$ (similar to the case of inferring $s$ from $f_{target}()$~\cite{Song2020Overlearning}).

\setlength\tabcolsep{1pt}
\begin{table}[h]
\begin{center}
\caption{Low Pearson Correlation of $s$ with $y$, $x$, $\phi(s)$ and $\phi(x)$ indicates that model is memorizing unintended private data. \label{tbl:correlation}}
\begin{tabular}{ |c|c|c|c|c|}
\hline
\textbf{Dataset} & \multicolumn{2}{c|}{\textbf{y}} & \multicolumn{2}{c|}{\textbf{x}}\\
 & \textbf{\race} & \textbf{\sex} & \textbf{\race} & \textbf{\sex} \\
\hline
\textbf{CENSUS} & 0.02 & 0.01 & 0.00 $\pm$ 0.02  & 0.00 $\pm$ 0.02 \\
\textbf{COMPAS} & -0.06 & 0.02 & -0.01 $\pm$ 0.03 & -0.02 $\pm$ 0.05 \\
\textbf{LAW}    & 0.02 & 0.02 &  -0.01 $\pm$ 0.02 &  0.00 $\pm$ 0.01 \\
\textbf{CREDIT} & 0.01 & -0.01 & 0.01 $\pm$ 0.01 & 0.00 $\pm$ 0.02 \\
\hline
\rowcolor{LightCyan} \multicolumn{5}{|c|}{\textbf{\intgrad}} \\
\hline
\textbf{Dataset} & \multicolumn{2}{c|}{\textbf{$\phi(s)$}} & \multicolumn{2}{c|}{\textbf{$\phi(x)$}} \\
 & \textbf{\race} & \textbf{\sex} & \textbf{\race} & \textbf{\sex} \\
\hline
\textbf{CENSUS} & 0.00 $\pm$ 0.02 & 0.00 $\pm$ 0.02 & 0.00 $\pm$ 0.02 & 0.00 $\pm$ 0.02\\
\textbf{COMPAS} & -0.01 $\pm$ 0.02 & -0.01 $\pm$ 0.06 &  -0.01 $\pm$ 0.02  & 0.00 $\pm$ 0.07\\
\textbf{LAW} & 0.02 $\pm$ 0.00 & 0.02 $\pm$ 0.02 & 0.02 $\pm$ 0.00 & 0.01 $\pm$ 0.02 \\
\textbf{CREDIT} & 0.01 $\pm$ 0.02 & 0.00 $\pm$ 0.03 & 0.01 $\pm$ 0.02  &  0.00 $\pm$ 0.02\\
 \hline
\rowcolor{LightCyan} \multicolumn{5}{|c|}{\textbf{\deeplift}} \\
\hline
\textbf{Dataset} & \multicolumn{2}{c|}{\textbf{$\phi(s)$}} & \multicolumn{2}{c|}{\textbf{$\phi(x)$}} \\
 & \textbf{\race} & \textbf{\sex} & \textbf{\race} & \textbf{\sex} \\
\hline
\textbf{CENSUS} & 0.00 $\pm$ 0.02 & 0.00 $\pm$ 0.02 & 0.00 $\pm$ 0.02  & 0.00 $\pm$ 0.02\\
\textbf{COMPAS} & 0.00 $\pm$ 0.03  & -0.02 $\pm$ 0.06 & -0.01 $\pm$ 0.02 & 0.01 $\pm$ 0.06 \\
\textbf{LAW} & -0.01 $\pm$ 0.03 & -0.00 $\pm$ 0.01 & -0.02 $\pm$ 0.03  &  0.00 $\pm$ 0.01\\
\textbf{CREDIT} & 0.00 $\pm$ 0.01 & 0.00 $\pm$ 0.01 &  0.00 $\pm$ 0.02 & 0.00 $\pm$ 0.02 \\
 \hline
\rowcolor{LightCyan} \multicolumn{5}{|c|}{\textbf{\gshap}} \\
\hline
\textbf{Dataset} & \multicolumn{2}{c|}{\textbf{$\phi(s)$}} & \multicolumn{2}{c|}{\textbf{$\phi(x)$}} \\
 & \textbf{\race} & \textbf{\sex} & \textbf{\race} & \textbf{\sex} \\
\hline
\textbf{CENSUS} & 0.00 $\pm$ 0.02 & 0.00 $\pm$ 0.02 & 0.00 $\pm$ 0.02 &  0.00 $\pm$ 0.02 \\
\textbf{COMPAS} & -0.01 $\pm$ 0.04 & -0.01 $\pm$ 0.03 &  -0.03 $\pm$ 0.02 &  -0.01 $\pm$ 0.03\\
\textbf{LAW}    & -0.01 $\pm$ 0.02 & 0.00 $\pm$ 0.00 &  -0.02 $\pm$ 0.00  & 0.00 $\pm$ 0.00\\
\textbf{CREDIT} & 0.00 $\pm$ 0.01 & 0.01 $\pm$ 0.02 &  0.00 $\pm$ 0.01 & 0.01 $\pm$ 0.02\\
 \hline
\rowcolor{LightCyan} \multicolumn{5}{|c|}{\textbf{\smgrad}} \\
\hline
\textbf{Dataset} & \multicolumn{2}{c|}{\textbf{$\phi(s)$}} & \multicolumn{2}{c|}{\textbf{$\phi(x)$}} \\
 & \textbf{\race} & \textbf{\sex} & \textbf{\race} & \textbf{\sex} \\
\hline
\textbf{CENSUS} & 0.00 $\pm$ 0.02 & 0.00 $\pm$ 0.02 & 0.00 $\pm$ 0.02 & 0.00 $\pm$ 0.02\\
\textbf{COMPAS} & 0.00 $\pm$ 0.02 & -0.01 $\pm$ 0.07 &  -0.01 $\pm$ 0.02  & 0.00 $\pm$ 0.08 \\
\textbf{LAW}    & -0.04 $\pm$ 0.04 & -0.01 $\pm$ 0.03 & -0.03 $\pm$ 0.06  &  0.01 $\pm$ 0.03 \\
\textbf{CREDIT} & 0.01 $\pm$ 0.02 & 0.00 $\pm$ 0.02 & 0.01 $\pm$ 0.02 & 0.00 $\pm$ 0.02 \\
 \hline
\end{tabular}
\end{center}
\end{table}

%% file: 5setup.tex
\section{Experimental Setup}\label{sec:setup}

We describe the tabular benchmark datasets used in our evaluation (Section~\ref{sec:datasets}), $f_{target}$ and $f_{adv}$ architectures (Section~\ref{sec:arch}), and metrics for evaluating attack success (Section~\ref{sec:metrics}).

\subsection{Datasets}\label{sec:datasets}

We consider four tabular datasets to demonstrate different high-stakes decision making applications. All the datasets have a binary classification task and publicly available.

\begin{itemize}[leftmargin=*]

\item \noindent\textbf{Adult Income (CENSUS)} comprises of 48,842 data records with 14 attributes about individuals from 1994 US Census data.
The attributes include marital status, education, occupation, job hours per week among others.
The binary classification is whether an individual makes an income of 50k per annum.

\item \noindent\textbf{Recidivism (COMPAS)} is used for commercial algorithms by judges and parole officers for estimating the likelihood of a criminal reoffending. It contains 10,000 criminal defendants in Florida.
The binary classification is if a criminal will reoffend.

\item \noindent\textbf{Law School Dataset (LAW)} is based on survey conducted by Law School Admission Council across 163 law schools in the United States. It contains information on 21,790 law students such as their entrance exam scores (LSAT), their grade-point average (GPA) collected prior to law school, and their first year average grade.
The classification is to predict if an applicant will have a high first year average grade.

\item \noindent\textbf{UCI Credit Card (CREDIT)} is an anonymized dataset from the UCI Machine Learning dataset repository and contains information about different credit card applicants. The dataset contains 30,000 records with 24 attributes for each record. The binary classification is if the application was approved.
\end{itemize}

In all the four datasets, the sensitive attributes are \race and \sex. We use this sensitive attributes for demonstration. The attack however can extend to other sensitive attributes with discrete values.
We use 70\% of $\mathcal{D}$ for $f_{target}$ and the remaining 30\% as testing dataset. $\mathcal{D}_{aux}$ is 50\% of the testing dataset while the other half is used as unseen dataset for evaluating the attack success. 


\subsection{Architecture}\label{sec:arch}

We now describe the model architectures and training hyperparameters for $f_{target}$, trained on the main classification task, and $f_{adv}$ used by \adv to map the explanations to $s$. We use pytorch and captum library for model explanations and our code is made publicly available: \url{https://github.com/vasishtduddu/AttInfExplanations.git}.

\begin{itemize}[leftmargin=*]

\item \noindent\textbf{Target Models.} We consider a fully connected neural network with four hidden layers of sizes [1024, 512, 256, 128] for all the datasets. Note that all the datasets have a binary classification tasks and the target models are binary classifiers. The target models are trained for 30 epochs with Adam optimizer and learning rate of 1e-3 and no regularization. 

\item \noindent\textbf{Attack Models.} We use a neural network model for all datasets other than LAW, where we use a random forest classifier with a maximum depth of 150. We consider a fully connected neural network with three hidden layers of sizes [64, 128, 32]. The model is trained using Adam optimizer with a learning rate of 1e-3 trained for 500 epochs.

The attack methodology is independent of ML models used and can be evaluated easily against other architectures.

\item \noindent\textbf{Model Accuracy.} The model utility is computed over the unseen test dataset across all the datasets. The test accuracy for the CENSUS dataset is 82.20\%, CREDIT dataset is 77.92\%, COMPAS dataset is 74.67\%, and LAW dataset is 95.63\%.

\end{itemize}

\subsection{Metrics}\label{sec:metrics}

We consider three main metrics for evaluating the success of attribute inference attack.
\begin{itemize}[leftmargin=*]

\item \noindent\underline{\textbf{Precision}}. The ratio of true positives to the sum of true positive and false positives. This indicates the fraction of $s$ inferred as having a positive value by \adv which indeed have positive attribute value as ground truth. 

\item \noindent\underline{\textbf{Recall}}. The ratio of true positives to the sum of true positives and false negatives. This indicates the fraction of $s$ with positive values which are correctly inferred by \adv.

\item \noindent\underline{\textbf{F1 Score.}} The harmonic mean of precision and recall computed as $2\times\frac{precision. recall}{precision + recall}$. The highest value is one indicating perfect precision and recall while the minimum value of zero, when either precision or recall is zero.
\end{itemize}

%% file: 7results_tm1.tex
\section{\ref{threatmodel1}: Evaluation of Attack Success}\label{sec:tm1results}

We first consider \ref{threatmodel1}, where \adv has access to $\phi(x \cup s)$ and hence $\phi(x)$ and $\phi(s)$. We evaluate attack success to infer $s$ from $\phi(x \cup s)$ (Section~\ref{subsec:worstcase}). Followed by this, we evaluate attack success to infer $s$ from only $\phi(s)$ (Section~\ref{subsec:correlation}).

\subsection{Inferring $s$ from $\phi(x \cup s)$}\label{subsec:worstcase}

We first evaluate the simplest attack surface: \adv has access to the entire model explanation vector $\phi(x \cup s)$ to infer $s$.
\adv's $f_{adv}$ maps the entire explanation to $s$, i.e., $f_{adv}: \phi(x \cup s) \rightarrow s$. Our hypothesis is that $\phi(x \cup s)$ is distinguishable for different values of $s$ which is captured by  $f_{adv}$.

\setlength\tabcolsep{5pt}
\begin{table}[!htb]
\caption{\underline{\ref{threatmodel1}: Inferring $s$ from $\phi(x \cup s)$.}}
\label{tab:worstcase}
\begin{center}
\begin{tabular}{ | c | c | c | c | } 
 \hline
\rowcolor{LightCyan}  \multicolumn{4}{|c|}{\textbf{\intgrad}} \\
  \hline
  \textbf{Dataset} & \textbf{Recall} & \textbf{Precision} & \textbf{F1-Score}\\ 
   & \race | \sex  & \race | \sex & \race | \sex \\
 \hline
 \textbf{\small CENSUS} & 0.98 | 0.95 & 0.95 | 0.87 & 0.97 | 0.91 \\
\textbf{\small COMPAS}  & 1.00 | 1.00 & 1.00 | 0.98 & 1.00 | 0.99 \\
\textbf{\small CREDIT}  &  0.95 | 0.94 &  0.69 | 0.61 & 0.80 | 0.74 \\
\textbf{\small LAW} & 0.93 | 0.57 & 0.92 | 0.55 & 0.93 | 0.56 \\
 \hline
\rowcolor{LightCyan} \multicolumn{4}{|c|}{\textbf{\deeplift}} \\
 \hline
 \textbf{Dataset} & \textbf{Recall} & \textbf{Precision} & \textbf{F1-Score}\\ 
  & \race | \sex  & \race | \sex & \race | \sex \\
   \hline
  \textbf{\small CENSUS} &  0.99 | 0.98  & 0.95 | 0.93  & 0.97 | 0.96 \\
   \textbf{\small COMPAS} & 1.00 | 0.97 & 1.00 | 0.98 & 1.00 | 0.97 \\
 \textbf{\small CREDIT} &  0.95 | 0.91  & 0.70 | 0.61 & 0.81 | 0.73 \\  
\textbf{\small LAW} & 0.94 | 0.88 & 0.92 | 0.54 & 0.93 | 0.67  \\
 \hline
\rowcolor{LightCyan} \multicolumn{4}{|c|}{\textbf{\gshap}} \\
 \hline
 \textbf{Dataset} & \textbf{Recall} & \textbf{Precision} & \textbf{F1-Score}\\ 
  & \race | \sex  & \race | \sex & \race | \sex \\
   \hline
 \textbf{\small CENSUS}&  0.94 | 0.97 & 0.98 | 0.94 & 0.96 | 0.95 \\
 \textbf{\small COMPAS} &  0.94 | 0.96 & 0.95 | 0.94 &  0.95 | 0.95 \\
 \textbf{\small CREDIT} &  0.97 | 0.95  & 0.70 | 0.61 & 0.81 | 0.74 \\
 \textbf{\small LAW} &  0.99 | 0.93 & 0.99 | 0.95 & 0.99 | 0.94 \\
\hline
\rowcolor{LightCyan} \multicolumn{4}{|c|}{\textbf{\smgrad}}\\
 \hline
 \textbf{Dataset} & \textbf{Recall} & \textbf{Precision} & \textbf{F1-Score}\\ 
  & \race | \sex  & \race | \sex & \race | \sex \\
   \hline
\textbf{\small CENSUS}&  0.97 | 0.95 & 0.94 | 0.90  & 0.95 | 0.93 \\ 
 \textbf{\small COMPAS} & 0.99 | 0.99 & 1.00 | 0.99 & 0.99 | 0.99  \\ 
\textbf{\small CREDIT} & 0.93 | 0.92  & 0.70 | 0.61  & 0.80 | 0.73  \\ 
\textbf{\small LAW} & 0.99 | 0.99 & 1.00 | 1.00  & 0.99 | 0.99 \\ 
\hline
\end{tabular}
\end{center}
\end{table}

As seen in Table~\ref{tab:worstcase}, we indeed validate our hypothesis. The attack success as measured using F1-Score are high: \intgrad (\sex: 0.80 $\pm$ 0.16; \race: 0.92 $\pm$ 0.07), \deeplift (\sex: 0.83 $\pm$ 0.13; \race: 0.92 $\pm$ 0.07), \gshap (\sex: 0.89 $\pm$ 0.08; \race: 0.92 $\pm$ 0.06) and \smgrad (\sex: 0.91 $\pm$ 0.10; \race: 0.91 $\pm$ 0.10). In addition to high F1-Scores, the high precision and recall values indicate that our proposed attribute inference attack is effective to infer $s$ from $\phi(x \cup s)$.

\subsection{Inferring $s$ from $\phi(s)$}\label{subsec:correlation}

We now consider a different setting for fine-grained analysis: can \adv exploit only $\phi(s)$ to infer $s$? Here, $\phi(s)$ is directly influenced by $s$ while $\phi(x)$ is indirectly influence by $s$.
Our hypothesis that $\phi(s)$ is sufficient for reasonable attack success and does not require entire model explanation $\phi(x \cup s)$ to successfully infer $s$. \adv's $f_{adv}$ infers $s$ using only its corresponding explanation, i.e., $f_{adv}: \phi(s) \rightarrow s$. 

\setlength\tabcolsep{5pt}
\begin{table}[!htb]
\caption{\underline{\ref{threatmodel1}: Inferring $s$ from $\phi(s)$.}}
\label{tab:attribute_correlation}
\begin{center}
\begin{tabular}{ | c | c | c | c | } 
 \hline
\rowcolor{LightCyan}  \multicolumn{4}{|c|}{\textbf{\intgrad}} \\
  \hline
  \textbf{Dataset} & \textbf{Recall} & \textbf{Precision} & \textbf{F1-Score}\\ 
   & \race | \sex  & \race | \sex & \race | \sex \\
 \hline
 \textbf{\small CENSUS} & 1.00 | 0.96 & 0.90 | 0.75  & 0.94 | 0.84 \\
 \textbf{\small COMPAS}  & 0.99 | 0.98 & 1.00 | 0.94 & 0.99 | 0.96 \\
 \textbf{\small CREDIT}  &  1.00 | 1.00 &  0.70 | 0.82 & 0.61 | 0.75 \\
 \textbf{\small LAW} & 0.98 | 1.00 & 1.00 | 1.00 & 0.98 | 1.00 \\
 \hline
\rowcolor{LightCyan} \multicolumn{4}{|c|}{\textbf{\deeplift}} \\
 \hline
 \textbf{Dataset} & \textbf{Recall} & \textbf{Precision} & \textbf{F1-Score}\\ 
  & \race | \sex  & \race | \sex & \race | \sex \\
   \hline
 \textbf{\small CENSUS}&  1.00 | 1.00 & 0.99 | 0.70 & 0.99 | 0.82 \\
\textbf{\small COMPAS} & 0.94 | 1.00 & 0.97 | 0.81 & 0.95 | 0.89 \\  
\textbf{\small CREDIT} & 1.00 | 0.99 & 0.70 | 0.61 & 0.82 | 0.75 \\
\textbf{\small LAW} & 0.60 | 1.00 & 0.99 |1.00 & 0.75 | 1.00 \\
 \hline
\rowcolor{LightCyan} \multicolumn{4}{|c|}{\textbf{\gshap}} \\
 \hline
 \textbf{Dataset} & \textbf{Recall} & \textbf{Precision} & \textbf{F1-Score}\\ 
  & \race | \sex  & \race | \sex & \race | \sex \\
   \hline
 \textbf{\small CENSUS} & 0.99 | 1.00  & 0.90 | 0.66  & 0.94 | 0.79 \\
\textbf{\small COMPAS} & 0.94 | 1.00 & 0.96 | 0.81  & 0.95 | 0.89 \\
\textbf{\small CREDIT} & 1.00 | 1.00 & 0.70 | 0.60 & 0.82 | 0.75 \\  
\textbf{\small LAW} & 0.99 | 0.99 & 0.93 | 0.55  & 0.96 | 0.71  \\
\hline
\rowcolor{LightCyan} \multicolumn{4}{|c|}{\textbf{\smgrad}}\\
 \hline
 \textbf{Dataset} & \textbf{Recall} & \textbf{Precision} & \textbf{F1-Score}\\ 
  & \race | \sex  & \race | \sex & \race | \sex \\
   \hline
\textbf{\small CENSUS}&  1.00 | 0.79 & 0.90 | 0.73  & 0.94 | 0.76 \\ 
\textbf{\small COMPAS}& 0.98 | 0.99 & 1.00 | 0.93 & 0.99 | 0.96 \\ 
\textbf{\small CREDIT} & 0.99 | 0.99 & 0.70 | 0.61  & 0.82 | 0.75  \\ 
\textbf{\small LAW} &  1.00 | 1.00 & 1.00 | 0.56 & 1.00 | 0.72 \\ 
\hline
\end{tabular}
\end{center}
\end{table}

We validate our hypothesis as indicated by the high F1-Score: \intgrad (\sex: 0.88 $\pm$ 0.09; \race: 0.88 $\pm$ 0.15), 
\deeplift (\sex:0.86 $\pm$ 0.09; \race: 0.87 $\pm$ 0.09),
\gshap(\sex: 0.78 $\pm$ 0.06; \race: 0.91 $\pm$ 0.05), and
\smgrad (\sex: 0.79 $\pm$ 0.09; \race: 0.93 $\pm$ 0.07). In addition to high F1-Scores, the high precision and recall values indicate that our proposed attribute inference attack is effective to infer $s$ from only $\phi(s)$.

\begin{remark}
In \ref{threatmodel1}, the high attack success is attributed to the distinguishability of model explanations for different values of $s$. In other words, different values of $s$ \textbf{explicitly} influence the model predictions as they are included in the training dataset. This in-turn results in distinguishable explanations for different values of $s$. This distinguishability is captured by training $f_{adv}$ to infer $s$.  
\end{remark}

%% file: 8results_tm2.tex
\section{\ref{threatmodel2}: Evaluation of Attack Success}\label{sec:tm2results}

Having shown that our proposed attack is successful in \ref{threatmodel1}, we now evaluate the attack success in \ref{threatmodel2}. We show the attack success on exploiting $\phi(x)$ (Section~\ref{subsec:explonly}) followed by exploiting combination of $f_{target}(x)$ and $\phi(x)$ (Section~\ref{subsec:combined}).

\subsection{Inferring $s$ from $\phi(x)$}\label{subsec:explonly}


We evaluate the effectiveness of our attack to exploit $\phi(x)$ which are the only explanations available to \adv.
\adv maps $\phi(x)$ to value of $s$ using the trained attack ML model, i.e., $f_{adv}: \phi(x) \rightarrow s$. Our hypothesis is that despite $s$ not directly being included in the training dataset and input, some attributes in $x$ might act as a proxy for $s$. Hence, $s$ influences model predictions \textit{indirectly} resulting in distinguishable model explanations for different values of $s$.

\setlength\tabcolsep{5pt}
\begin{table}[!htb]
\caption{\underline{\ref{threatmodel2}: Inferring $s$ from $\phi(x)$.}}
\label{tab:explanations}
\begin{center}
\begin{tabular}{ | c | c | c | c | } 
 \hline
\rowcolor{LightCyan}  \multicolumn{4}{|c|}{\textbf{\intgrad}} \\
  \hline
  \textbf{Dataset} & \textbf{Recall} & \textbf{Precision} & \textbf{F1-Score}\\ 
   & \race | \sex  & \race | \sex & \race | \sex \\
 \hline
 \textbf{\small CENSUS} & 0.97 | 0.85 & 0.90 | 0.79 & 0.94 | 0.82 \\ 
\textbf{\small COMPAS}  & 0.76 | 0.99  & 0.57 | 0.80  & 0.65 | 0.89 \\ 
 \textbf{\small CREDIT}  & 0.91 | 0.91 &  0.69 | 0.60 & 0.79 | 0.72 \\
 \textbf{\small LAW} & 0.98 | 0.90  & 0.94 | 0.56  & 0.96 | 0.69 \\
 \hline
\rowcolor{LightCyan} \multicolumn{4}{|c|}{\textbf{\deeplift}} \\
 \hline
 \textbf{Dataset} & \textbf{Recall} & \textbf{Precision} & \textbf{F1-Score}\\ 
  & \race | \sex  & \race | \sex & \race | \sex \\
   \hline
 \textbf{\small CENSUS} &  0.98 | 0.90  & 0.91 | 0.80  & 0.94 | 0.85 \\
 \textbf{\small COMPAS} & 0.81 | 1.00 & 0.54 | 0.81  & 0.65 | 0.89 \\
 \textbf{\small CREDIT} & 0.98 | 0.91 & 0.70 | 0.60 & 0.81 | 0.72 \\
 \textbf{\small LAW} & 0.99 | 0.99  & 0.92 | 0.55  & 0.96 | 0.70  \\
 \hline
\rowcolor{LightCyan} \multicolumn{4}{|c|}{\textbf{\gshap}} \\
 \hline
 \textbf{Dataset} & \textbf{Recall} & \textbf{Precision} & \textbf{F1-Score}\\ 
  & \race | \sex  & \race | \sex & \race | \sex \\
   \hline
 \textbf{\small CENSUS} & 0.94 | 0.85 & 0.90 | 0.80 & 0.92 | 0.83 \\
 \textbf{\small COMPAS} & 0.75 | 0.90 & 0.55 | 0.82 & 0.63 | 0.86 \\
 \textbf{\small CREDIT} & 0.95 |  0.95 & 0.70 | 0.61 & 0.80 | 0.74  \\
 \textbf{\small LAW} & 0.93 | 0.53 & 0.92 | 0.55 & 0.93 | 0.54 \\
\hline
\rowcolor{LightCyan} \multicolumn{4}{|c|}{\textbf{\smgrad}}\\
 \hline
 \textbf{Dataset} & \textbf{Recall} & \textbf{Precision} & \textbf{F1-Score}\\ 
  & \race | \sex  & \race | \sex & \race | \sex \\
   \hline
 \textbf{\small CENSUS} & 0.98 | 0.87 & 0.90 | 0.78  & 0.94 | 0.82  \\
\textbf{\small COMPAS} & 0.77 | 0.98 & 0.56 | 0.80  & 0.65 | 0.89\\ 
\textbf{\small CREDIT} & 0.92 | 0.88  & 0.70 | 0.60 & 0.79 | 0.72 \\ 
\textbf{\small LAW}  & 0.97 | 0.96 & 0.94  | 0.55 &  0.96 | 0.70\\ 
\hline
\end{tabular}
\end{center}
\end{table}

We confirm this hypothesis in Table~\ref{tab:explanations} which indicates high attack success. For instance, F1-Score across four datasets for each explanation algorithm are as follows: \intgrad (\sex: 0.78 $\pm$ 0.07; \race: 0.83 $\pm$ 0.12), \deeplift (\sex: 0.79 $\pm$ 0.08; \race: 0.84$\pm$ 0.12), \gshap (\sex: 0.74 $\pm$ 0.12 \race: 0.82 $\pm$ 0.12), and \smgrad (\sex: 0.78 $\pm$ 0.07; \race: 0.83 $\pm$ 0.12). Hence, censoring $s$ is ineffective to mitigate privacy risk to attribute inference attacks.

\subsection{Inferring $s$ from $f_{target}(x) \cup \phi(x)$}\label{subsec:combined}

Having shown the attack success on exploiting $\phi(x)$, we answer \textit{how good are explanations + predictions combination as an attack surface for \adv to exploit?}
We want to evaluate the impact on attack success on combining $f_{target}(x)$ with $\phi(x)$.

\setlength\tabcolsep{5pt}
\begin{table}[!htb]
\caption{\underline{\ref{threatmodel2}: Inferring $s$ from $f_{target}(x) \cup \phi(x)$.}}
\label{tab:combined}
\begin{center}
\begin{tabular}{ | c | c | c | c | } 
 \hline
\rowcolor{LightCyan}  \multicolumn{4}{|c|}{\textbf{\intgrad}} \\
  \hline
  \textbf{Dataset} & \textbf{Recall} & \textbf{Precision} & \textbf{F1-Score}\\ 
   & \race | \sex  & \race | \sex & \race | \sex \\
 \hline
 \textbf{\small CENSUS} & 0.99 | 0.72 & 0.90 | 0.66 & 0.94 | 0.69 \\
\textbf{\small COMPAS}  & 0.76 | 0.99 & 0.47 | 0.82 & 0.58 | 0.90 \\
\textbf{\small CREDIT}  &  0.89 | 0.90 & 0.70 | 0.60 & 0.78 | 0.72 \\
\textbf{\small LAW} & 0.98 | 0.98 & 0.93 | 0.55  & 0.95 | 0.70 \\ 
 \hline
\rowcolor{LightCyan} \multicolumn{4}{|c|}{\textbf{\deeplift}} \\
 \hline
 \textbf{Dataset} & \textbf{Recall} & \textbf{Precision} & \textbf{F1-Score}\\ 
  & \race | \sex  & \race | \sex & \race | \sex \\
   \hline
 \textbf{\small CENSUS} &  0.99 | 0.75 & 0.90 | 0.66  & 0.94 | 0.70\\
 \textbf{\small COMPAS} &  0.75 | 0.99 & 0.49 | 0.81 & 0.59 | 0.89 \\
 \textbf{\small CREDIT}  &  0.97 | 0.92 & 0.80 | 0.60 & 0.81 | 0.73 \\
 \textbf{\small LAW} & 0.99 | 0.99 & 0.92 | 0.54 & 0.95 | 0.70 \\
 \hline
\rowcolor{LightCyan} \multicolumn{4}{|c|}{\textbf{\gshap}} \\
 \hline
 \textbf{Dataset} & \textbf{Recall} & \textbf{Precision} & \textbf{F1-Score}\\ 
  & \race | \sex  & \race | \sex & \race | \sex \\
   \hline
   \textbf{\small CENSUS} & 0.95 | 0.60 & 0.90 | 0.66 & 0.93 | 0.63 \\
\textbf{\small COMPAS}  & 0.55 | 0.93 & 0.50 | 0.81 & 0.52 | 0.86 \\
\textbf{\small CREDIT} & 0.93 | 0.92 & 0.69 | 0.61 & 0.79 | 0.73 \\
\textbf{\small LAW} &  0.83 | 0.58 & 0.92 | 0.55 & 0.87 | 0.56 \\
\hline
\rowcolor{LightCyan} \multicolumn{4}{|c|}{\textbf{\smgrad}}\\
 \hline
 \textbf{Dataset} & \textbf{Recall} & \textbf{Precision} & \textbf{F1-Score}\\ 
  & \race | \sex  & \race | \sex & \race | \sex \\
   \hline
 \textbf{\small CENSUS} &  0.90 | 0.72 & 0.90 | 0.66  & 0.90 | 0.69 \\
\textbf{\small COMPAS} &  0.84 | 0.88 & 0.69 | 0.61 & 0.76 | 0.72 \\ 
  \textbf{\small CREDIT} & 0.69 | 0.99  & 0.47 | 0.81 & 0.56 | 0.89 \\ 
\textbf{\small LAW} & 0.97 | 0.95 & 0.92 | 0.54 & 0.95 | 0.69\\ 
\hline
\end{tabular}
\end{center}
\end{table}

Given the combination $f_{target}(x) \cup \phi(x)$ as input, \adv trains $f_{adv}$ to map it to $s$, i.e.,  $f_{adv}: (f_{target}(x) \cup \phi(x)) \rightarrow s$.
In Table~\ref{tab:combined}, we note that the attack success does not show a significant difference compared to the results in Table~\ref{tab:explanations} for exploiting only model explanations. 
Furthermore, for \race, the attack success degrades compared to using only model explanations (Table~\ref{tab:explanations}). Here, we conjecture that the model predictions lower the distinguishability for $f_{adv}$ to infer $s$ compared to only using model explanations.
These observations indicate that model explanations are a strong attack surface for \adv to exploit independent of model predictions.

\begin{remark}
In \ref{threatmodel2}, similar to \ref{threatmodel1}, the high attack success is attributed to the distinguishability of model explanations for different values of $s$. Unlike \ref{threatmodel1}, different values of $s$ \textbf{implicitly} influence the model predictions via other attributes acting as proxy variables for $s$. This in-turn results in distinguishable explanations for different values of $s$ which is exploited by $f_{adv}$.  
\end{remark}

%% file: 9predatt_compare.tex
\section{Comparing Privacy Risk of Explanations vs. Predictions}\label{sec:compare}

Having shown the success of attack on model explanations, we answer \textit{how risky are explanations compared to model predictions with respect to attribute inference attacks?}
The experimental setup in our work is the same as Aalmoes et al.~\cite{aalmoes2022dikaios}.  
Hence, we report the results from Aalmoes et al.~\cite{aalmoes2022dikaios} as the state-of-the-art for attribute inference attacks. Specifically, we consider their \pr attack for both \ref{threatmodel1} and \ref{threatmodel2}. 

\setlength\tabcolsep{1.75pt}
\begin{table}[!htb]
\caption{Reported state-of-the-art attribute inference attack success exploiting model predictions from Aalmoes et al.~\cite{aalmoes2022dikaios}.}
\label{tab:prediction}
\begin{center}
\begin{tabular}{ | c | c | c | c | }
\hline
\rowcolor{LightCyan} \multicolumn{4}{|c|}{\textbf{\pr Attack (w/o $S$)}} \\
\hline
\multirow{2}{*}{\textbf{Dataset}}  & \textbf{Recall} & \textbf{Precision} & \textbf{F1-Score}  \\
& \race | \sex & \race | \sex & \race | \sex   \\
\hline
\textbf{CENSUS} &  0.91 | 0.94 & 0.90 | 0.69 & 0.90 | 0.80  \\
\textbf{COMPAS}  &  0.97 | 0.96 &  0.48 | 0.82 & 0.64 | 0.88  \\
\textbf{LAW} & 0.98 | 1.00 & 0.95 | 0.56 & 0.96 | 0.72 \\
\textbf{CREDIT}  & 0.99 | 0.97  & 0.69 | 0.61 & 0.81 | 0.75  \\
\hline
 \hline
 \rowcolor{LightCyan}  \multicolumn{4}{|c|}{\textbf{\pr Attack (w/ $S$)}} \\
 \hline
\multirow{2}{*}{\textbf{Dataset}} & \textbf{Recall} & \textbf{Precision} & \textbf{F1-Score}  \\
& \race | \sex & \race | \sex & \race | \sex  \\
 \hline
 \textbf{CENSUS}  & 0.90 | 0.91  & 0.92 | 0.70 & 0.91 | 0.79  \\
 \textbf{COMPAS}   & 0.72 | 0.97 & 0.67 | 0.82 & 0.69 | 0.89  \\
\textbf{LAW}  & 0.98 | 0.96 & 0.97 | 0.57 & 0.97 | 0.72 \\
\textbf{CREDIT}  & 0.99 | 0.84 & 0.69 | 0.67 & 0.81 | 0.75 \\
 \hline
\end{tabular}
\end{center}
\end{table}

We compare the inference capability of $s$ from $\phi(x \cup s)$ (reported Table~\ref{tab:worstcase}) against the inference capability of using model predictions (reported Table~\ref{tab:prediction} (w/ $s$)).
We note that the attack success in Table~\ref{tab:worstcase} for model explanations is higher than model predictions in Table~\ref{tab:prediction} in most of the cases. 
Similarly, when $s$ is not included in training data, we find that the performance reported in Table~\ref{tab:explanations} is better than the results in Table~\ref{tab:prediction} (w/o $s$).

In summary, model explanations alone are stronger attack surface for attribute inference attack compared to model predictions. 

%% file: 10related.tex
\section{Related Work}\label{sec:related}

We discuss some prior works which have indicated security and privacy vulnerabilities for model explanations.

\noindent\textbf{Security Attacks on Model Explanations.} Model explanations are sensitive to distribution shifts and adversarial examples. Model explanations do not accurately reflect the biases in ML model leading to misleading explanations which influence user trust in black box models~\cite{10.1145/3375627.3375833}.
Adversarial examples can be generated for model misclassification as well as fooling interpretations~\cite{undefire,heo2019fooling}. The attack exploits the fact that model predictions and their interpretations are misaligned.
SHAP and LIME explanations algorithms have also been shown to be vulnerable to adversarial examples~\cite{slack2020fooling,slack2021feature}.
Counterfactual examples are an alternative approach for explanations which are not robust: they converge to different counterfactuals under a small perturbation~\cite{slack2021counterfactual}.
To address these, Lakkaraju et al.~\cite{lakkaraju2020robust} propose adversarial training with minimax objective to construct high fidelity explanations with respect to the worst-case adversarial perturbations.
Additionally, Yeh et al.~\cite{yeh2019fidelity} propose two measures for evaluating robustness of explanations: sensitivity and infidelity, and propose algorithms to improve both.

\noindent\textbf{Privacy Attacks on Model Explanations.} Prior works have indicated a trade-off between transparency and privacy.
Model explanations have been shown to be vulnerable to membership inference attacks where \adv aims to infer whether a given data record belonged to the model training data using model explanations~\cite{shokri2021aies}. This threat was extended to data reconstruction attacks for explanations which reveal training data instances. To incorporate membership privacy and transparency, model explanations with differential privacy have been proposed in literature~\cite{patel2020model,harder2020interpretable}. However, this comes at the cost of quality of explanations.
Furthermore, since model explanations characterize the model's decision boundary, it can be used to steal the functionality of a model using model extraction attacks~\cite{10.1145/3287560.3287562,aivodji2020model}.
None of the prior works evaluate the vulnerability to attribute inference attacks.

%% file: 11conclusions.tex
\section{Discussions and Conclusions}\label{sec:conclusions}

\noindent\textbf{Summary.} Model explanations assign scores to attributes of an input by estimating their influence to model prediction. These model explanations potentially leak sensitive attributes. We propose the first attribute inference attack on model explanations and show their effectiveness in two threat models. 
We show yet another trade-off between privacy and transparency in ML models.

\noindent\textbf{Attribute Privacy Risk Metric.} There is a need to design data privacy risk assessment tools as required by several privacy laws such as GDPR (Article 35). However, there is limited prior work on estimating privacy risk of different sensitive attributes to inference attacks: Hannun et al.~\cite{hannun2021fil} propose a generic metric based on Fisher Information Loss which are shown to estimate privacy risk to attribute inference attacks. However, it is applicable only to linear and convex models and hence, not scalable to deep neural networks with non-linear and non-convex objective. Furthermore, they limit \adv to unbiased estimators which they indicate will be violated in the presence of $\mathcal{D}_{aux}$. 

We discuss the viability of model explanations as a tool for attribute privacy risk assessment. We indicate different requirements to be satisfied for attribute privacy risk metric and indicate how model explanations satisfy them.
\begin{enumerate}[leftmargin=*]
    \item \textit{Independent of Attacks.} The metric should estimate the attribute privacy risk scores \textit{without} using any specific attacks. The assigned scores should capture the root cause of attribute privacy risk, i.e., different values of $s$ have different influence on model prediction which can be exploited by \adv to infer the value of $s$. This makes the privacy risk scores to quantify privacy risk to all possible future attacks. 
    \begin{itemize}[leftmargin=*]
        \item Model explanations are independent of any specific attribute inference attacks and capture the influence of attributes to the model predictions.
    \end{itemize}
    \item \textit{Correlation with Attacks.} The attribute privacy risk scores assigned to each record's sensitive attribute should correlate with the attack success to infer $s$. This ensures that the privacy risk scores capture the susceptibility to attack success.
    \begin{itemize}[leftmargin=*]
    \item Model explanations can be mapped to $s$ as shown in this work (Section~\ref{subsec:correlation}) which can allow for model explanations as a relative privacy risk measure.
    \end{itemize}
    \item \textit{Efficient and Scalable.} The computation of scores should be efficient and scale to large deep neural network architectures.
    \begin{itemize}[leftmargin=*]
    \item Model explanations can be efficiently computed on deep neural networks and scalable to large models. 
    \end{itemize}
\end{enumerate}

We leave the careful design and evaluation of attribute privacy risk metric based on model explanations for future work.

\noindent\textbf{Defences Against Attribute Inference Attacks.} Current literature lacks specific defences against the described attribute inference attacks as well as prior attacks leveraging model predictions. AttriGuard was proposed as a method to lower the success of \adv's attack ML model by adding adversarial noise to \adv's auxiliary data obtained from public sources~\cite{attriguard}. This defence is more generic and can be adapted to ML models: \mb can use vulnerability of model explanations to adversarial examples, proposed in prior literature~\cite{slack2020fooling,slack2021feature,undefire,heo2019fooling,10.1145/3375627.3375833,yeh2019fidelity,slack2021counterfactual,lakkaraju2020robust}, as a defence mechanism to lower the success of \adv's attack model. 
Data sanitization to remove the privacy risk while maintaining the utility of the ML model have also been explored~\cite{dysan}. Finallly, model explanations with differential privacy~\cite{patel2020model,harder2020interpretable} can possibly lower the privacy risk to attribute inference attacks as the minimize the influence of individual data records as a whole. However, using mechanisms based on pufferfish privacy~\cite{pufferfish,pufferfishmechanisms,zhang2022attribute} is likely address attribute inference risks. However, these have not been explored in the context of model explanations.
We keep the evaluation of defences for future work.

\noindent\textbf{Algorithmic Fairness and Attack Success.} There are several algorithms which guarantee fairness across sensitive attributes in ML models~\cite{fair1,fair2,fair3}. It is unclear whether there is a correlation between model bias and attack success to infer $s$ from model explanations. We speculate that since many evaluated datasets have proxy attributes to $s$, attribute inference attacks might still be effective (see Section~\ref{subsec:explonly}). A detailed study of the impact of algorithmic fairness on attribute inference attacks of $s$ is left for future work.

\section*{Acknowledgement}

The first author was supported in part by Intel (in the context of the Private-AI Institute).